\documentclass[aip,amsmath,amssymb,reprint,apl]{revtex4-1}

\usepackage{graphicx}
\usepackage{dcolumn}
\usepackage{bm}

\usepackage[utf8]{inputenc}
\usepackage[T1]{fontenc}
\usepackage{mathptmx}
\usepackage{etoolbox}

\makeatletter
\def\@email#1#2{%
 \endgroup
 \patchcmd{\titleblock@produce}
  {\frontmatter@RRAPformat}
  {\frontmatter@RRAPformat{\produce@RRAP{*#1\href{mailto:#2}{#2}}}\frontmatter@RRAPformat}
  {}{}
}
\makeatother

\begin{document}
\title{Spin Hall Nano-Oscillator Empirical Electrical Model for Optimal On-chip Detector Design}

\author{Rafaella Fiorelli}
\email{fiorelli@imse-cnm.csic.es}
 \affiliation{Instituto de Microelectrónica de Sevilla, CSIC and Universidad de Sevilla, Sevilla, 41092.}

\author{Mona Rajabali}
 \affiliation{NanOsc AB, Kista, 16440 Sweden.}

\author{Roberto Méndez-Romero}
\affiliation{Instituto de Microelectrónica de Sevilla, CSIC and Universidad de Sevilla, Sevilla, 41092.}

\author{Akash Kumar}
\affiliation{Applied Spintronics Group, Department of Physics, University of Gothenburg, 41296 Gothenburg, Sweden.}
\affiliation{Research Institute of Electrical Communication and Center for Science and Innovation in Spintronics, Tohoku University, 2-1-1 Katahira, Aoba-ku, Sendai 980-8577 Japan.}

\author{Artem Litvinenko}
\affiliation{Applied Spintronics Group, Department of Physics, University of Gothenburg, 41296 Gothenburg, Sweden.}

\author{Teresa Serrano-Gotarredona}
\affiliation{Instituto de Microelectrónica de Sevilla, CSIC and Universidad de Sevilla, Sevilla, 41092.}

\author{Farshad Moradi}
\affiliation{Integrated Circuits and Electronics Lab (ICELab), Electrical and Computer Engineering Department, Aarhus University, 8200 Aarhus N, Denmark.}

\author{Johan Åkerman}
\affiliation{Applied Spintronics Group, Department of Physics, University of Gothenburg, 41296 Gothenburg, Sweden.}
\affiliation{Research Institute of Electrical Communication and Center for Science and Innovation in Spintronics, Tohoku University, 2-1-1 Katahira, Aoba-ku, Sendai 980-8577 Japan.}

\author{Bernabé Linares-Barranco}
 \affiliation{Instituto de Microelectrónica de Sevilla, CSIC and Universidad de Sevilla, Sevilla, 41092.}

\author{Eduardo Peralías}
\affiliation{Instituto de Microelectrónica de Sevilla, CSIC and Universidad de Sevilla, Sevilla, 41092.}

\begin{abstract}
 As nascent nonlinear oscillators, nano-constriction spin Hall nano-oscillators (SHNOs) represent a promising potential for integration into more complicated systems such as neural networks, magnetic field sensors, and radio frequency (RF) signal classification, their tunable high-frequency operating regime, easy synchronization, and CMOS compatibility can streamline the process. To implement SHNOs in any of these networks, the electrical features of a single device are needed before designing the signal detection CMOS circuitry. This study centers on presenting an empirical electrical model of the SHNO based on a comprehensive characterization of the output impedance of a single SHNO, and its available output power in the range of 2-10 GHz at various bias currents. 
\end{abstract}

\maketitle

\section{Introduction}
\label{sec:introduction}

As complementary metal-oxide-semiconductor (CMOS) technology is reaching its physical limit, other technologies are about to thrive. In particular, nonlinear spiking and oscillatory spintronic devices exhibit tremendous potential in leading-edge areas such as emulating the spiking behaviors of neurons \cite{Torrejon2017, Romera2018, Zahedinejad2020,houshang2022prappl,kumar2023mutual}, RF signal classification \cite{ross2023multilayer}, ultra-fast microwave spectral analysis \cite{litvinenko2022ultrafast}, and the possibility of implementing highly accelerated neuromorphic computing systems \cite{Müller2022,gonzalez2024spintronic}. 
Among the family of spintronic microwave oscillators, nano-constriction spin Hall nano-oscillators (SHNOs) are simple heavy metal/ferromagnet bilayers (HM/FM) through which the pure spin current is produced by passing DC bias current ($I_B$) and leads to a steady-state precession, known as auto-oscillation~\cite{demidov2014nanoconstriction,Chen2016,behera2024ultra}. Considering their facile fabrication, broad frequency tunability \cite{Zahedinejad2018}, easy injection locking\cite{Rajabali2023}, robust mutual synchronization \cite{Kumar2023}, and individual tunability \cite{Muralidhar2022,Khademi2023Large}, the SHNOs are particularly promising for various applications from
magnetic field sensors \cite{xie2023nanoscale} to neural network integration \cite{sethi2023compensation,kumar2024spin,gonzalez2024spintronic}.

When implementing an SHNO-based network, it is essential to ascertain the oscillation status of the SHNO (\textit{e.g.,} firing/non-firing in case of a neural network). However, due to the relatively low power and high noise of the output oscillating signal~\cite{litvinenko2023phase}, the CMOS detector and the SHNO necessitate achieving a good impedance matching \cite{Bendjeddou2023}. If not, reflecting a significant share of the available SHNO output power makes it more challenging to detect the SHNO signal, as confirmed in \cite{Fiorelli2023}.

In the realm of SHNO output signal detection, fully integrated approaches not only reduce detector size and power consumption but also are more adaptable for the joint integration of SHNO and detectors; whether homogeneous or heterogeneous. The selection of the optimal technology and detector's architecture is dependent on the SHNO's electrical characteristics, which include its working frequency, output power, noise, and output impedance \cite{Fiorelli2023}. Hence, an accurate electrical model of the SHNO is mandatory. This model must include information about the SHNO: (i) output power, (ii) noise levels, and (iii) output impedance. This information allows for knowing the SHNO signal-to-noise (SNR) ratio and therefore deduce the maximum accepted noise level of the receiver so that the SHNO signal is not masked by its noise. Moreover, an optimized design of the electrical circuitry that senses the SHNO output signal requires 
information on the SHNO output impedance to address an excellent impedance coupling between this detector and the SHNO, minimizing the power losses. 

It is known that the SHNO output power increases as its oscillation frequency goes up, which in turn, allows a quicker classification of its state \cite{fulara2019spin}. But this increase in frequency must be accompanied by a guarantee of observation of the oscillation by the integrated detector, whose implementation is limited to its (i) maximum detection frequency and, (ii) minimum practicable bandwidth (BW) associated with the SNR. In this study, we demonstrate that excellent trade-offs in terms of SHNO output power, detector's feasibility, and performance, can be achieved by selecting the optimal oscillating frequency around 6~GHz. Furthermore, the empirical electrical model of the SHNO is extracted and presented for the frequency range of 2-10 GHz.

The SNR will not be notably favorable when the detector possesses a broad BW \cite{Pozar2011} (\textit{i.e.,} exceeding 100 MHz as noted later in Section III-A), which is likely to be the case in fully integrated detectors. Hence, achieving the desired signal matching and developing an accurate electrical AC SHNO model is critical if we aim to leverage the full potential of the available SNR.
 
Although other types of nano-oscillators were modeled electrically \cite{Bendjeddou2023}, there is a notable absence of an AC SHNO model that comprehensively describes its output impedance, power characteristics, and noise equivalent output. For the first time, this paper introduces an AC empirical electrical model for an SHNO derived from experimental measurements. This AC evaluation holds immense significance since only considering the DC range characterization fails to provide sufficient insights for an optimal signal detector design, as in \cite{Fiorelli2023}. This AC empirical model is developed by the data captured from two sets of experiments that measure the SHNO load output power, $P_L$, and the output impedance, $Z_{SHNO}$. 
 
\begin{figure}
\centerline{\includegraphics[width=1\columnwidth, angle=0]{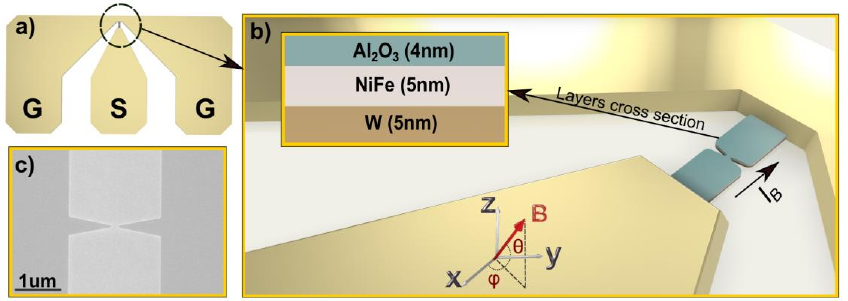}}
\caption{(a) Fabricated SHNO with GSG CPW. (b) Magnified scheme of SHNO geometry and the contact pads. The inset displays the detail of the heterostructure on which SHNO is fabricated. (c) The SEM top-view image of a single 180 nm wide SHNO.}
\label{fig:SHNO_Struct_Model}
\end{figure}
 
This paper is outlined as follows. Section II presents fabrication details and the experimental setup to measure the SHNO impedance and power. The measurement results are displayed in Section III, whereas the empirical electrical model is deployed in Section IV. Finally, Section~V concludes our approach to develop this empirical electrical model for SHNOs for an optimal on-chip detector design.

\section{Experimental part} \label{sec:experiment}

The 180-nm wide SHNOs, used in the measurements to extract the empirical electrical model, were fabricated on heterostructures consisting of W(5nm)/Py(5nm)(Ni$_{80}$Fe$_{20}$)/Al$_{2}$O$_{3}$ (4nm) \cite{Kumar2023},\cite{litvinenko2023phase}. The fabrication process involves DC/RF magnetron sputtering of the stack at room temperature onto a high-resistance silicon substrate (20$\times$20 mm$^2$) with a base pressure of less than 3$\times$10$^{-8}$ Torr. The heterostructure is then patterned into 8$\times$12 $\mu$m$^{2}$ rectangles with bow-tie shaped NCs using a Raith EBPG 5200 electron beam lithography (EBL) system followed by Ar-ion etching. Optical lithography is used to define the ground-signal-ground (GSG) coplanar waveguide (CPW), which is followed by a deposition and liftoff process using a bi-layer of Cu(500nm)/Pt(20nm) as top contacts ({Fig.}~\ref{fig:SHNO_Struct_Model}(a)). A magnified scheme of the SHNO is presented in Fig.~\ref{fig:SHNO_Struct_Model}(b), along with the SHNO layer arrangement details in the cross-sectional view (inset). Furthermore, a top-view scanning electron microscopy (SEM) image of the fabricated SHNO is seen in Fig.~\ref{fig:SHNO_Struct_Model}(c) [more details of fabrication can be found in \cite{kumar2022fabrication}].

\begin{figure}
\centerline{\includegraphics[width=1\columnwidth]{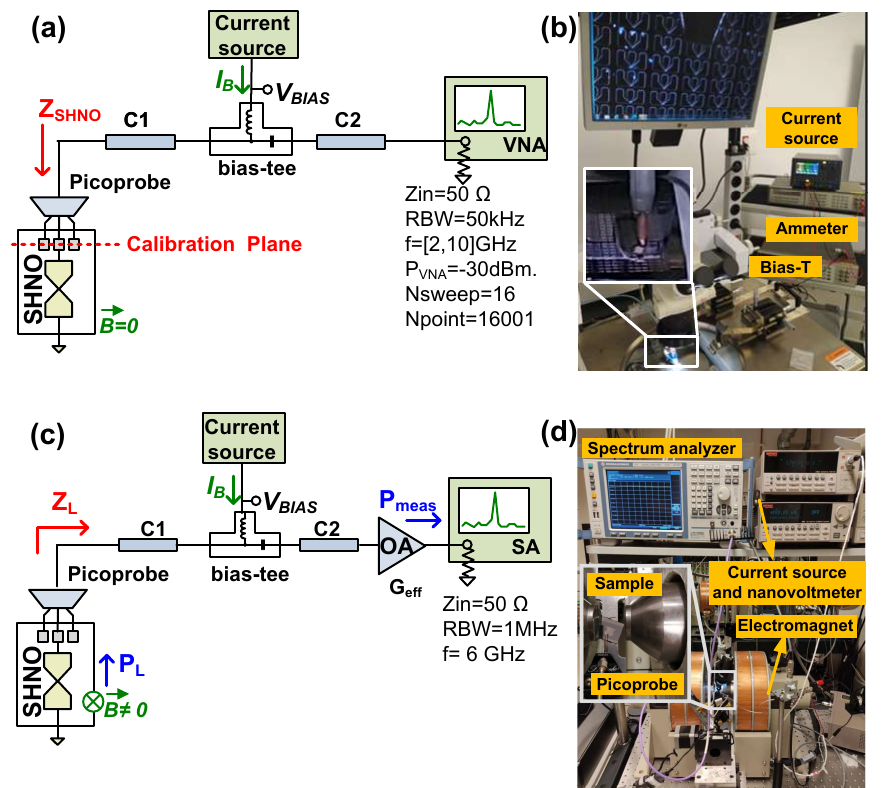}}
\caption{ Schematic and setup photo for Experiment 1 (a-b) and Experiment 2 (c-d).}
\label{fig:Experiment_1_2}
\end{figure}

\subsection{Impedance setup}

\textit{Experiment 1} focuses on the impedance measurement, with the corresponding setup scheme and photo presented in Fig.~\ref{fig:Experiment_1_2}(a) and (b), respectively. This measurement is conducted using an Agilent N5230A Vector Network Analyzer (VNA). The SHNO is biased via the Mini-circuits ZX85-12G+ bias-tee. Its output is connected to the VNA and evaluated between 2-10~GHz, under the values of $I_B=\{2.0, 2.25, 2.35, 2.45\}~mA$. To do impedance measurements over a biased device-under-test (DUT), all external electromagnetic signals of the DUT are voided except the one that injects into the DUT node where the impedance is to be known. Hence, this test is performed in the absence of an external magnetic field. Furthermore, although the frequency of interest to obtain the model is around 6~GHz, the wide range of frequencies of [2,10]~GHz was evaluated to assess its behavior, ensuring both a smooth behavior at high frequencies and a tendency to the expected values at the DC level.

The accuracy of the impedance measurement relies on the careful calibration of the entire setup to remove the effect of cables C1, C2, bias-tee, and picoprobe (Fig.~\ref{fig:Experiment_1_2}(a)). We made use of the Cascade Microtech GSG 150~$\mu$m-pitch P/N 101-190 impedance standard substrate to carry out this calibration, obtaining excellent results with a measured voltage standing wave ratio lower than 1.1 at [2,10]~GHz range. 

\subsection{Power and noise setup}

\textit{Experiment 2} aims to find the maximum SHNO output power ( $P_{meas\_peak}$) around 6~GHz, using a spectrum analyzer. Its setup and photo are given in Fig.~\ref{fig:Experiment_1_2}(c) and (d), respectively. The GSG picoprobe (GGB Industries) conveys the input $I_B$ and the output generated signal between the SHNO and a bias-tee (MITEQ BT4000). The auto-oscillation is driven by $I_B$ running to the sample through the bias-tee DC port. Then, the generated SHNO RF signal goes to a low noise amplifier (LNA, B\&Z BZ0218A) with a transducer power gain $G_{T}=23~dB$ and eventually reaches the Spectrum Analyzer (SA, R\&S FSV 40 GHz), where the spectra data is captured.

The optimum magnetic field $\overrightarrow{B}$ is obtained by preliminary field scans at several values of $I_B$. At the chosen magnetic field, the signal will then be recorded while increasing $I_B$ to reach a maximum power value such that a further increase in $I_B$ does not substantially increase the measured power, $P_{meas}$. 
Additionally, the second outcome of the experiment pertains to the noise floor level, which will be useful in further calculations. As a result, the sample is subjected to an external magnetic field ($|B|=0.68~T$) at out-of-plane ($\theta=84^{\circ}$) and in-plane ($\varphi=22^{\circ}$) angles. The value $I_B=2.45~mA$ above which $P_{meas}$ does not vary substantially is considered the one that generates $P_{meas\_peak}$. 
 
\section{Results and discussion}
\label{sec:results}

\subsection{Impedance results}

\begin{figure}[b!]
\centerline{\includegraphics[width=1\columnwidth]{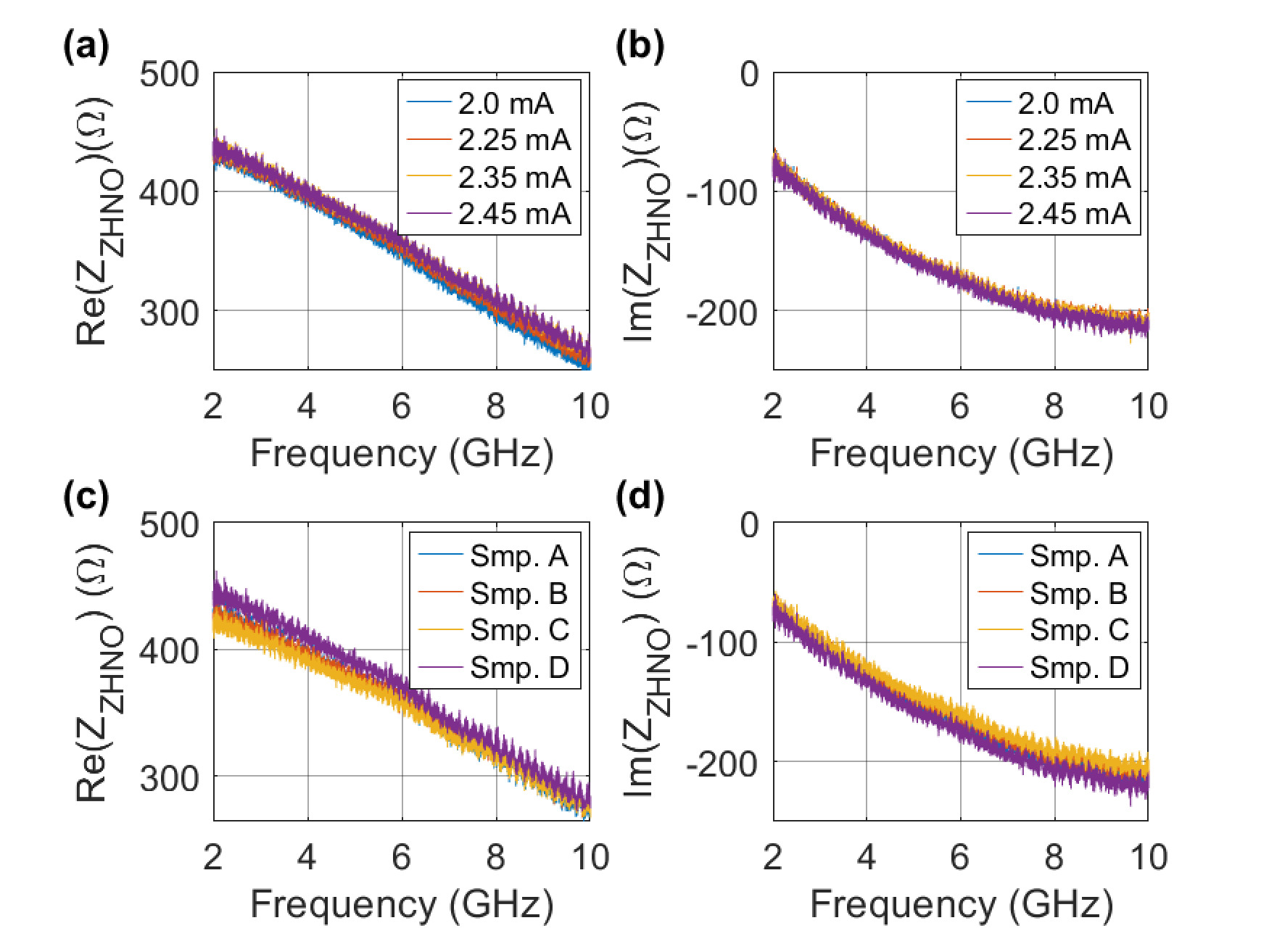}}
\caption{ Real and imaginary part of $Z_{SHNO}$ for four $I_B$ (a,b), and four SHNO samples at $I_B = 2.45~mA$ (c,d).}
\label{fig:Impedance}
\end{figure}

Impedance measurement results are deployed in Fig.~\ref{fig:Impedance}. In the RF/MW frequencies, the real part of $Z_{SHNO}$, $Re(Z_{SHNO})$, decreases almost linearly with frequency, reaching approximately 350~$\Omega$ at 6~GHz, and 250~$\Omega$ at 10~GHz (Fig.~\ref{fig:Impedance}(a)). This result is promising since making the SHNO work at higher frequencies implies a lower $Re(Z_{SHNO})$, simplifying the detector design and allowing it to reach optimal matching with the detector. In addition, Fig.~\ref{fig:Impedance}(a) verifies that $Re(Z_{{SHNO}})$ increases as $I_{B}$ increases.

 Fig.~\ref{fig:Impedance}(b) depicts the imaginary part $Z_{SHNO}$, $Im(Z_{SHNO})$, which diminishes as $I_B$ decreases for the whole frequency range. In particular, we observe a capacitive behavior over the entire frequency span under study. 

Finally, we measure the values of $Re(Z_{SHNO})$ and $Im(Z_{SHNO})$ for four SHNO identical samples to evaluate its dispersion. The findings, shown in Fig.~\ref{fig:Impedance}(c) and (d), yield a variation of 5\% and 2.5\% around 6~GHz, respectively.

\subsection{Power and noise results}

\begin{figure}
 \centering
\includegraphics[width=0.9\columnwidth]{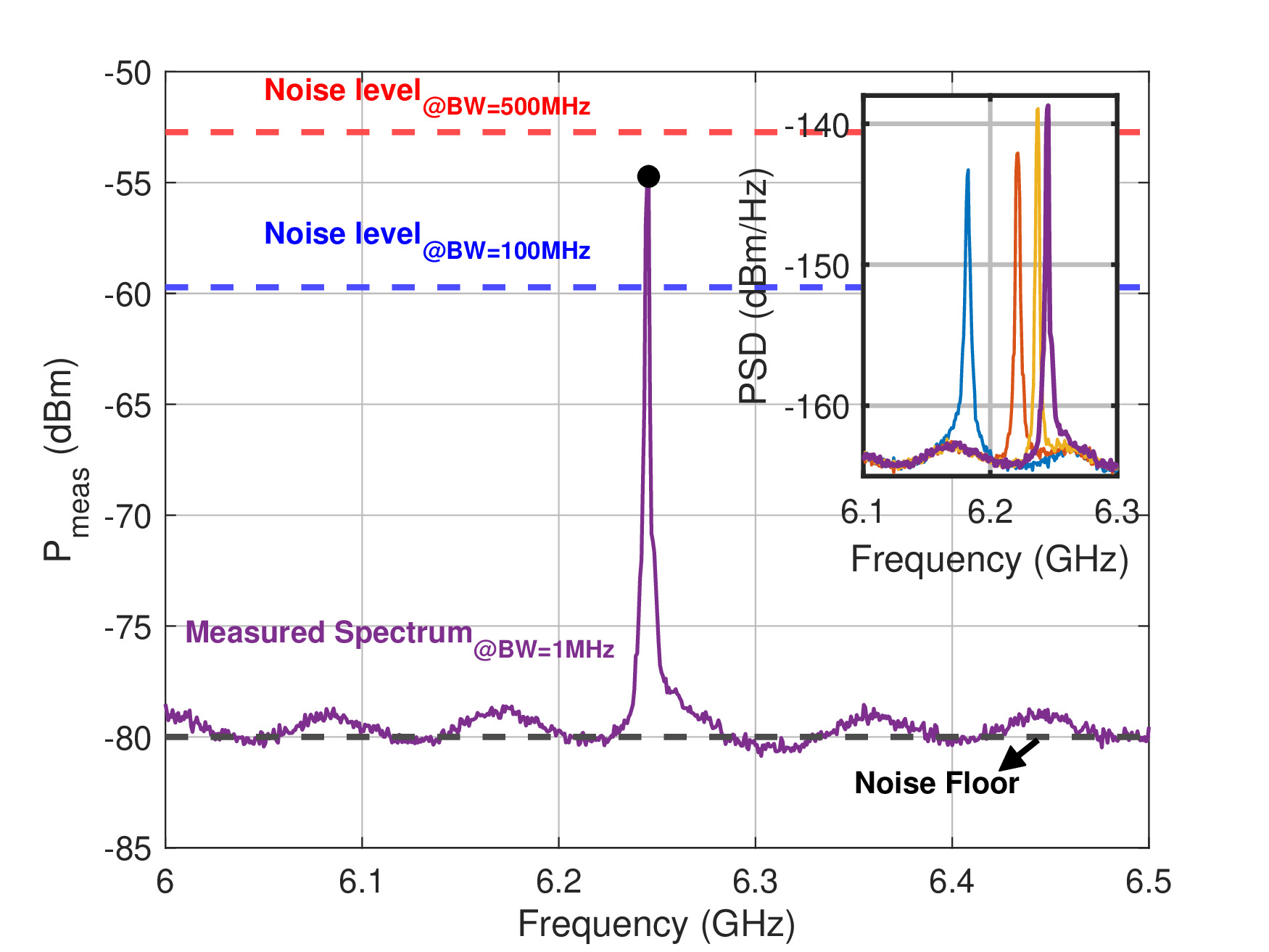}
 \caption{SHNO output spectrum for $I_B=2.45~mA$ with BW=1~MHz, considering noise levels associated with a detector with BW = [1, 100, 500] MHz. Inset: PSD for $I_B$ = [$2.0~mA$ (blue), $2.25~mA$ (orange), $2.35~mA$ (yellow), $2.45~mA$ (purple)].}
 \label{fig:Spectrum}
\end{figure}

The captured spectrum corresponding to $I_B= 2.45~mA$ is illustrated in Fig.~\ref{fig:Spectrum}, in the range of [6.0,6.5]~GHz. Using a BW of 1 MHz, the maximum measured power is $P_{meas \_peak}=-55~dBm$ at 6.25~GHz. This plot also yields noise floor power in the order of $P_{meas \_noise\_floor} \approx -80~dBm$. Then, an estimation of SHNO output signal power will be about $P_L \approx P_{meas \_peak} - G_{T} = -79~dBm$, and an estimation of the power spectral density (PSD) of the signal at the SHNO output is shown in the inset of Fig.~\ref{fig:Spectrum} with a noise floor power density of $P_{noise \_floor} \approx P_{meas \_noise \_floor} - 10~log_{10}(BW) - G_{T} = -163~dBm/Hz$.

The feasibility of the detection strongly depends on the detector BW, which is reflected in the SNR; for instance, SNR=$\{25dB_{@1MHz}, 5dB_{@100MHz}, -2dB_{@500MHz}\}$. Given that the BW is anticipated to be no less than 100 MHz in a monolithically integrated CMOS detector, we expect a low peak PSD level. This low SNR necessitates maximizing the available power at the detector input, highlighting the critical role of impedance matching.

\section{Electrical model} \label{sec:Model}

\begin{figure}[t!]
\centering

 \includegraphics[width=1\columnwidth]{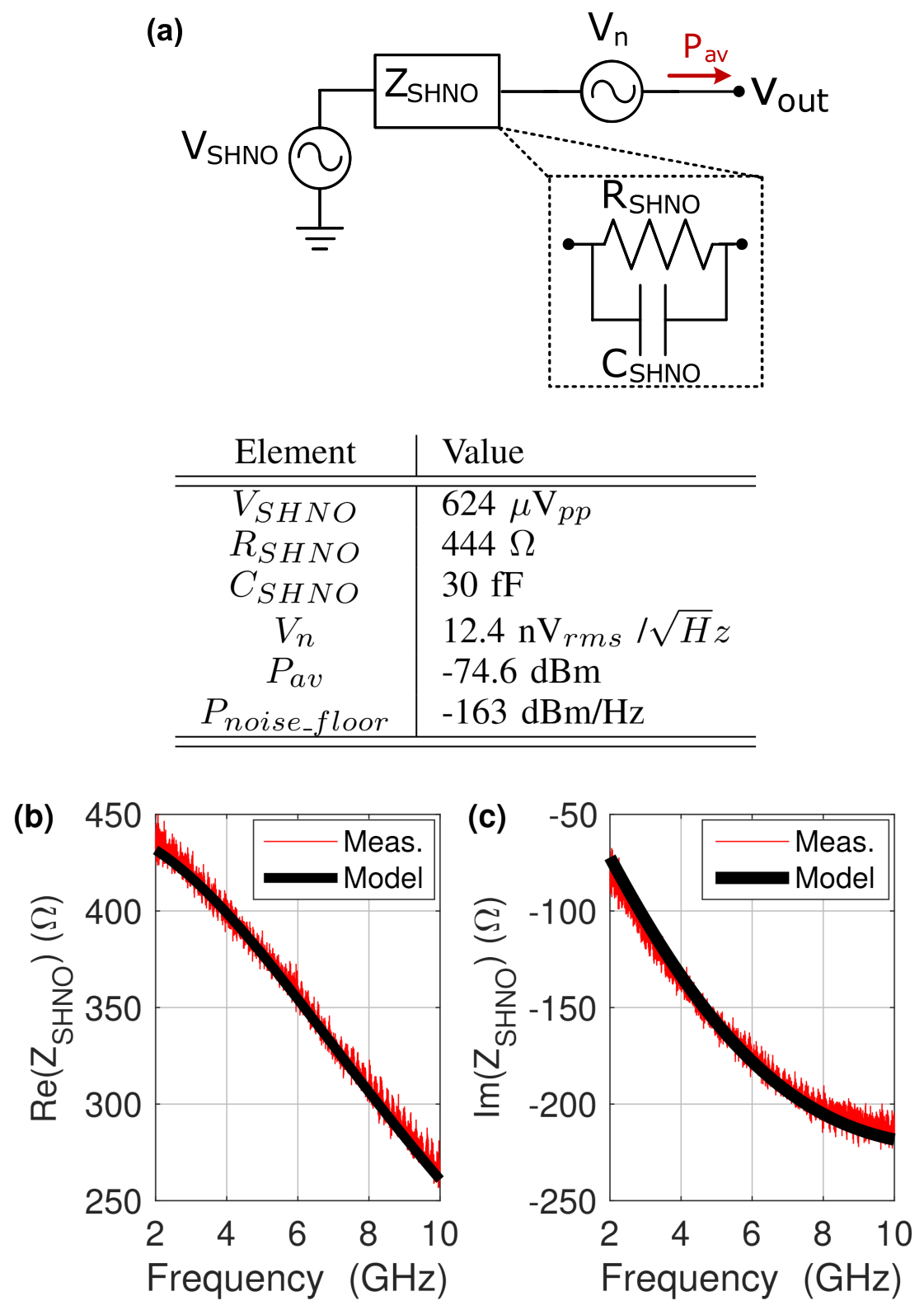}
\hspace{0.03cm}
\caption{(a) SHNO empirical electrical
model. (b) Real and (c) imaginary part of $Z_{SHNO}$ measured at $I_B=2.45mA$ (red line) and $Z_{SHNO}(\omega)^{model}$ (black).}
\label{fig:Impedance_Zoom}
\end{figure}

The information collected with \textit{Experiment 1} and \textit{Experiment 2} allows us to develop the SHNO Thévenin model, as seen in Fig.~\ref{fig:Impedance_Zoom}(a). It comprises the Thévenin voltage, $V_{SHNO}$, the equivalent output noise voltage $V_n$, and the equivalent output impedance $Z_{SHNO}$. The equations that describe this model. presented next, are derived from classical circuits' theory \cite{Alexander2016}.
The amplitude of the Thévenin's voltage, $|V_{SHNO}|$ and $Z_{SHNO}$, are related to the estimated output peak power, $P_L (\approx -79dBm)$, over the load impedance $Z_L$, by the following expression: 
\begin{equation}\label{eq:V_shno}
 \left |V_{SHNO}\right |= \left| Z_{SHNO}+Z_L\right| \sqrt{\frac{2P_L }{Re(Z_L)}}
 \end{equation} 
where $Z_L$ is the impedance seen at the output of the SHNO of the experimental setup shown in Fig.~\ref{fig:Experiment_1_2}(c), ideally, $Z_L=50 \Omega$. 

The proposed impedance model is presented in the inset of Fig.~\ref{fig:Impedance_Zoom}(a), being consistent with the device impedance measurements in DC. It comprises a resistor $R_{SHNO}$ in parallel with a capacitor $C_{SHNO}$, and where
\begin{equation}\label{eq:Zshnp}
Z_{SHNO}(\omega)^{model}=\frac{1}{1/R_{SHNO}+j\omega C_{SHNO}}
\end{equation}

The values of $R_{SHNO}$ and $C_{SHNO}$ are listed in the table embedded in Fig.~\ref{fig:Impedance_Zoom}(a), which are valid over the whole frequency range of [2,10] GHz (for $I_B= 2.45~mA$). This is reflected in Fig.~\ref{fig:Impedance_Zoom}(b-c), where the real and imaginary parts of $Z_{SHNO}$ are correctly described over the whole frequency span with the black curves of $Z_{SHNO}(\omega)^{model}$.

The root-mean-square of noise voltage source, $v_n$, is obtained by using \eqref{eq:V_shno}, and substituting $2P_L$ for the estimated floor noise power, $P_{{noise \_floor}} (\approx-163 dBm/Hz)$. 

 Finally, to complete the empirical model, we bring the available power gain of the SHNO, $P_{av,SHNO}$. As it is known, this is the maximum power of the SHNO that can be delivered to the load, and it occurs when the load is conjugately matched to the SHNO output impedance. It does not depend on the load and gives the designer a maximum value of power that could be delivered to the load, although this is never feasible in practice. This expression is,

\begin{equation} \label{eq:Pav}
 P_{av,SHNO} = \frac{|V_{SHNO}|^2}{8Re(Z_{SHNO})}
\end{equation}
 
 The numerical values of the empirical electrical model are listed in the table of Fig.~\ref{fig:Impedance_Zoom}(a). They have been obtained from the experimental quantities presented in subsections IIIA and IIIB and the expressions (1)-(3).

To conclude this section, we provide a practical example of how critical is to provide a good SHNO electrical model in designing an associated signal detector with an input impedance $Z_{det}$. Assuming the SHNO device interfaces with a noiseless detector featuring a practical BW of several hundred MHz, the detector output is anticipated to exhibit a maximum SNR of around 3 dB (Refer to Fig.~\ref{fig:Spectrum}).

\begin{figure}
\centering
\includegraphics[width=0.9\columnwidth]{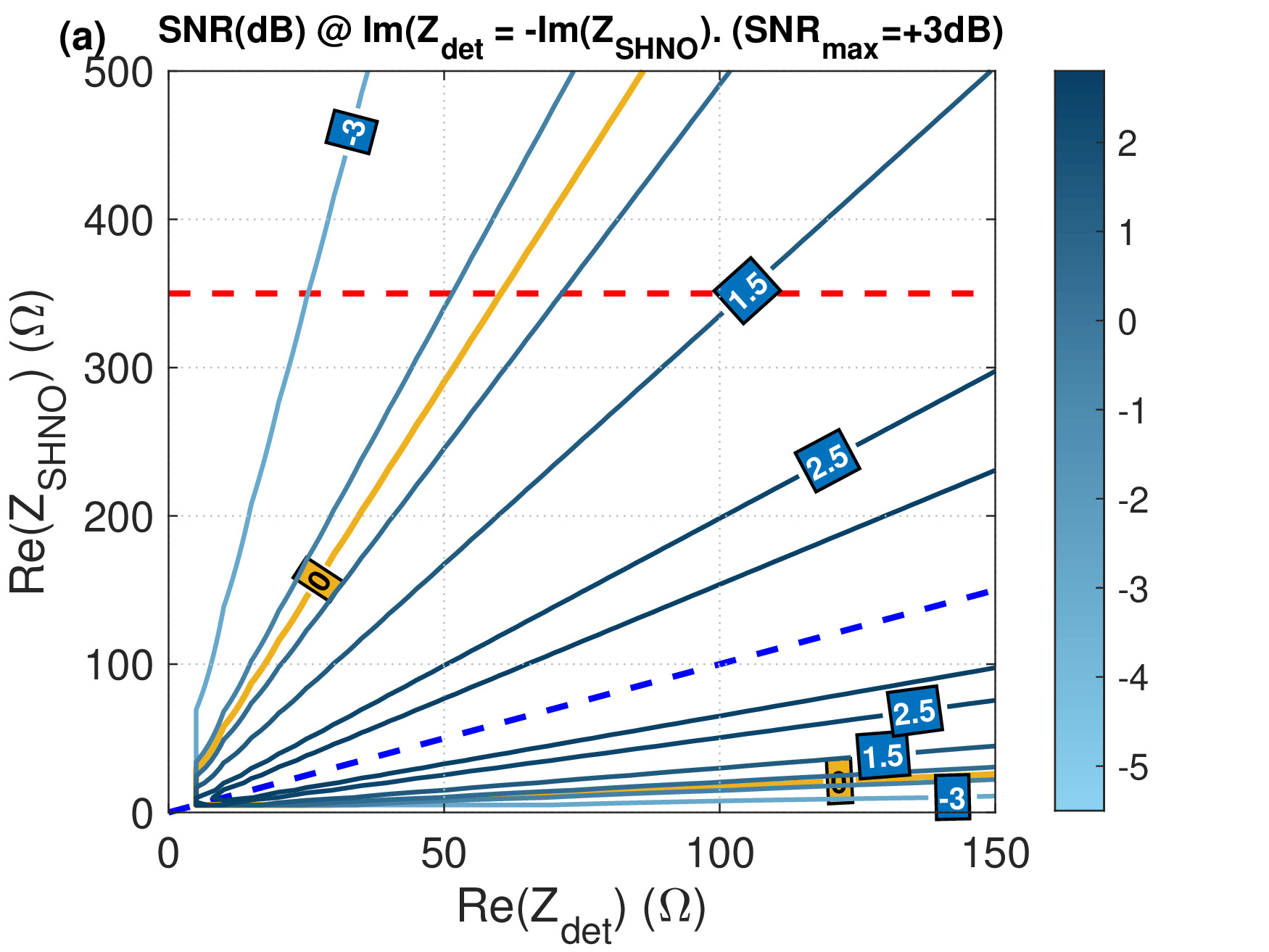} 
\includegraphics[width=0.9\columnwidth]{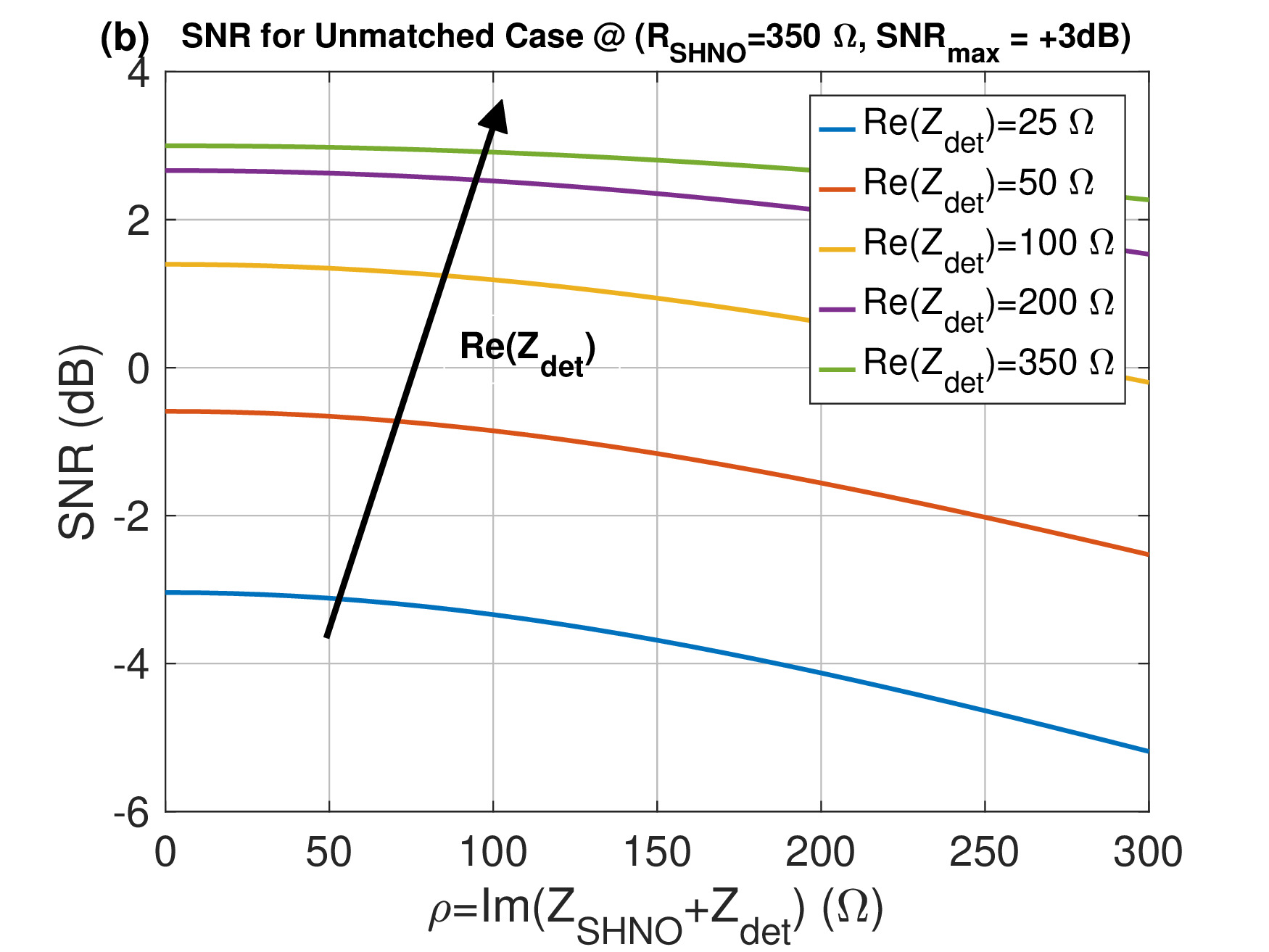} 
\hspace{0.03cm}
\caption{SNR at the input of the detector: (a) versus the real parts of $Z_{det}$ and $Z_{SHNO}$ impedances with opposite imaginary parts, and (b) the imaginary part of $Z_{SHNO}+Z_{det}$ for different $Re(Z_{det})$ at $R_{SHNO}=350\Omega$.}
\label{fig:Example_maps}
\end{figure}

Owing to the high value of $R_{SHNO}( \approx 350 \Omega$ at 6~GHz), the main challenge of achieving the matching condition lies in equalizing $Re({Z_{det}})$ to $R_{SHNO}$. This is particularly troublesome when attempting on-chip detectors, to achieve a joint integration of multiple units. To illustrate this challenge, \cite{Fiorelli2023} has reported the difficulty of designing on-chip detectors in CMOS technologies at 4.7~GHz, with $Re(Z_{det})$ above 100~$\Omega$.

Presuming the input impedance of the detector, $Z_{det}$, can be complex, and considering the maximum power-transfer theorem, the seek of maximum power transfer from the SHNO to the detector implies that $Z_{det}=Z_{SHNO}^*$. This is especially desirable due to the very low value of $P_{av,SHNO}$.

To elaborate on the effect of the mismatch between the SHNO and the detector in the SNR of the system, with a maximum value of SNR = +3 dB, one can initially consider a partial conjugation such that only the imaginary components of the impedance are matched, i.e. $Im(Z_{det})=-Im(Z_{shno})$. When both $Re(Z_{det})$ and $Re(Z_{SHNO})$ depart from equality, the SNR falls below 0 (see Fig.~\ref{fig:Example_maps}(a)). For example, at $Re(Z_{SHNO})=350 \Omega$ (red dash line), for $Re(Z_{det})<100\Omega$, the SNR is below 1.5~dB, and thus eliminating most of the margin that guarantees the detection of the oscillation signal.

Let's now consider the case when the net imaginary component, $\rho=|Im(Z_{SHNO}+Z_{det})|$ departs from the ideal condition ($\rho=0$). For $\rho=$[0,300]$~\Omega$ and $Re(Z_{det})$=[25,350] ~$\Omega$ we obtain the plot shown in Fig.~\ref{fig:Example_maps}(b). For instance, when $\rho>$100~$\Omega$ and $Re(Z_{det})<$ 200~$\Omega$, the deviation from matching conditions causes SNR to drop below 0~dB and therefore making it impossible to detect the oscillation signal. In other words, to ensure signal detection, it's crucial to design the detector with the goal of minimizing the net imaginary component, ideally approaching $\rho=0$, while simultaneously optimizing $Re(Z_{det})$ to closely match the oscillator resistance. By establishing these key criteria, we proposed a practical solution for on-chip detection of the SHNO signal, offering insights for future implementations of this nonlinear oscillator in complex networks.

\section{Conclusions}

This paper presents an electrical empirical model of the SHNO working in the 6-GHz range, which is drawn from a comprehensive study of the SHNO output impedance and its output power and noise levels shown at the SHNO signal detector. From the results of the study, especially due to the high $Re(Z_{SHNO})$ values and the non-negligible capacitive effect, it is clear that there is a need to provide an empirical electrical model to the designer of the fully integrated detector, and thus make the discrimination of the SHNO operating state feasible.

\section*{Acknowledgement}
This work was supported in part by the Horizon 2020 Research and Innovation Program No.
899559 “SpinAge”, DOI 10.3030/899559.

\bibliography{Main.bib}

\end{document}